\documentclass[conference]{IEEEtran}
\IEEEoverridecommandlockouts
\usepackage{cite}
\usepackage{url}
\usepackage{amsmath,amssymb,amsfonts}
\usepackage{algorithmic}
\usepackage{graphicx}
\usepackage{textcomp}
\usepackage{xcolor}
\def\BibTeX{{\rm B\kern-.05em{\sc i\kern-.025em b}\kern-.08em
    T\kern-.1667em\lower.7ex\hbox{E}\kern-.125emX}}
\begin{document}

\title{Data-Driven Modeling of Seasonal Dengue Dynamics in Bangladesh: A Bayesian-Stochastic Approach
}

\author{\IEEEauthorblockN{Mahmudul Bari Hridoy}
\IEEEauthorblockA{\textit{Department of Mathematics \& Statistics} \\
\textit{Texas Tech University}\\
Lubbock, Texas, U.S.A. \\
bari.hridoy@ttu.edu}
\and
\IEEEauthorblockN{S M Mustaquim}
\IEEEauthorblockA{\textit{Department of Mathematical Sciences} \\
\textit{The University of Texas at El Paso}\\
El Paso, Texas, U.S.A. \\
smustaquim@miners.utep.edu}}

\maketitle

\begin{abstract}

Bangladesh's worsening dengue crisis, fueled by its tropical climate, poor waste management infrastructure, rapid urbanization, and dense population, has led to increasingly deadly outbreaks, posing a significant public health threat. To address this, we propose a nonlinear, time-nonhomogeneous SEIR model incorporating seasonality through a novel transmission rate function. The model parameters are estimated using Bayesian inference with the Metropolis-Hastings algorithm in a Markov Chain Monte Carlo (MCMC) framework, calibrated with real-life dengue data from Bangladesh. To account for stochasticity and better assess outbreak probabilities, we extend the model to a time-nonhomogeneous continuous-time Markov chain (CTMC) framework. Our model provides new insights that can guide policymakers and offer a robust mathematical framework to better combat this crisis.

\end{abstract}


\section{Introduction}

Dengue fever is a rapidly spreading mosquito-borne infectious disease caused by the dengue virus, primarily transmitted by Aedes aegypti mosquitoes. It is endemic in the tropical and subtropical regions of South and Southeast Asia, including countries like Bangladesh, and accounts for nearly three-quarters of the global dengue burden \cite{hossain2023twenty}. Recognized by the  World Health Organization (WHO) as a top ten global health threat \cite{who2019threats}, dengue has been assigned the highest emergency response grade, G3, due to its severe impact \cite{who2024dengue}. Dengue infection ranges from asymptomatic to severe, potentially fatal disease, with no specific treatment, requiring supportive care for mild cases and hospitalization for severe ones \cite{aguiar2022mathematical, cdc_dengueControl}. The virus exists in four serotypes (DENV-1 to DENV-4), causing up to 400 million infections annually \cite{cdc_dengue}, with 100 million cases leading to illness and 40,000 global deaths \cite{who2023dengue}.

Dengue fever was first recognized in Bangladesh during a 1964 outbreak, when it was referred to as ``Dacca fever'' \cite{aziz1967dacca}. After becoming sporadic since the 1960s, a major epidemic in 2000 firmly established the virus in the country \cite{yunus2001dengue}. In recent years, Bangladesh has faced increasingly severe dengue outbreaks, with the DENV-2 and DENV-3 variants predominating and two of the deadliest outbreaks occurring in 2019 and 2022 \cite{mahmud2024alarming}. The situation has deteriorated further in 2023, with 321,179 cases and 1,705 deaths recorded over the year, highlighting the escalating severity of the dengue crisis in the country \cite{DGHS2023}. Bangladesh is particularly vulnerable to dengue outbreaks due to its tropical and subtropical climate, characterized by heavy rainfall and high humidity, which create an ideal environment for Aedes mosquitoes. 
\footnotetext{© 2024 IEEE. Personal use of this material is permitted. Permission from IEEE must be obtained for all other uses, in any current or future media, including reprinting/republishing this material for advertising or promotional purposes, creating new collective works, for resale or redistribution to servers or lists, or reuse of any copyrighted component of this work in other works.
}

Numerous mathematical models have been developed to study the epidemiology of dengue, with some models incorporating detailed vector dynamics \cite{aguiar2022mathematical, ESTEVA1998131, feng1997competitive, nipa2021effect, zheng2018modelling}, while others simplifying the process by focusing on the host population alone \cite{aguiar2008epidemiology, anderson1992infectious, asaduzzaman2024analysis, perkins2006patterns, riad2021risk}. 
Although these models have been widely applied to dengue data, the effectiveness of various models in capturing outbreak dynamics, particularly in regions with distinct seasonal variations, remains underexplored.  

In this manuscript, we develop a nonlinear, time-nonhomogeneous SEIR (Susceptible-Exposed-Infectious-Recovered) model incorporating seasonality in transmission rates, driven primarily by rainfall, the key factor in seasonal dengue transmission in Bangladesh \cite{hasan2024two, rahman2020association}. 
Simple SEIR models can effectively capture disease transmission dynamics without the complexity of vector-host models, which often add unnecessary detail without improving accuracy \cite{pandey2013comparing}. Given the limited effectiveness of current dengue control measures, a straightforward and broadly accessible model is crucial \cite{aguiar2022mathematical}. To further understand the impact on disease outbreak probabilities, we extend the deterministic model to a time-nonhomogeneous continuous-time Markov chain (CTMC) model. We estimate the model parameters through a Bayesian inference approach, employing the Metropolis-Hastings algorithm within the Markov Chain Monte Carlo (MCMC) framework. This approach utilizes daily infectious data \cite{DGHS2023} from January 1, 2023, to December 31, 2023, to estimate the model parameters accurately. In addition, we compute the basic and seasonal reproduction numbers, as well as the probability of disease outbreaks, to analyze the model outcomes. The methodology, including data, parameterization, and analysis of reproduction numbers and outbreak probabilities, is detailed in Section \ref{sec:Methodology}. Numerical results and interpretations are in Section \ref{sec:Numerical Results}, with our overarching message in Section \ref{sec:Conclusion}.


\section{Methodology}
\label{sec:Methodology}

\subsection{Data Source}

The data used in this study is sourced from the official Dengue Dashboard of the Directorate General of Health Services (DGHS2023), Bangladesh \cite{DGHS2023}. This platform monitors real-time dengue data, providing daily updates on infections, hospitalizations, and deaths reported by hospitals and health centers nationwide. 

\subsection{Model Formulation}

\begin{figure}[htbp]
\centering
\includegraphics[width=.8\linewidth]{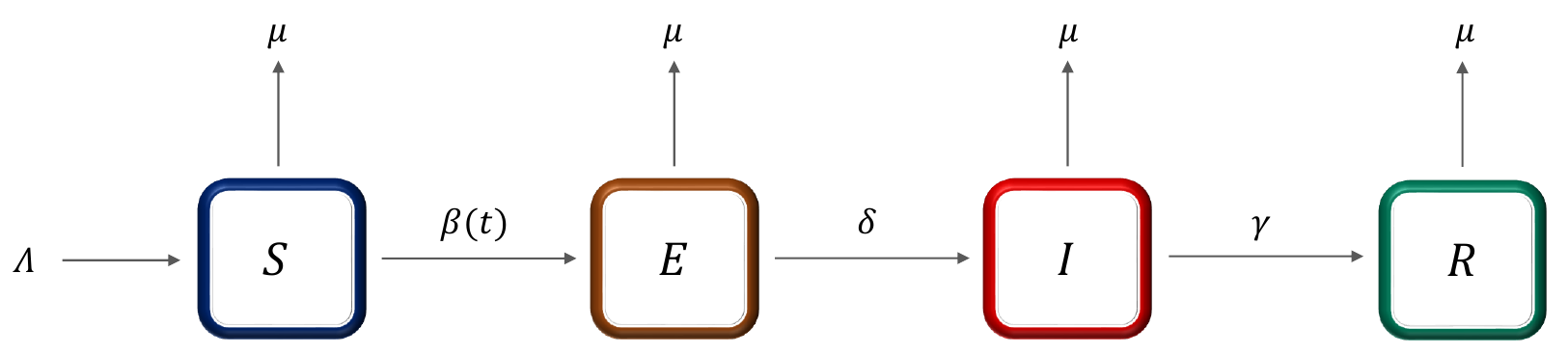} 
\caption{\label{fig:compartment} Compartmental diagram for the model.}
\end{figure}

The seasonal SEIR 
model parameters include the recruitment or birth rate $\Lambda=\mu N$, incubation rate $\delta$, natural death rate $\mu$, recovery rate $\gamma$, and seasonal transmission rate $\beta(t)$. $\beta(t)$ is a positive, time-dependent, periodic function with period $\omega > 0$ such that $\beta(t) = \beta(t + \omega)$. At any given time $t$, the total population is $N(t)= S(t)+E(t)+I(t)+R(t)$. The system of ODEs governing the model is presented in \eqref{eq:odeseir} and the compartmental diagram is illustrated in Fig.~\ref{fig:compartment}.

\begin{equation} \label{eq:odeseir}
\begin{cases}
\dot{S} &= \Lambda - \beta(t) \frac{I}{N} S - \mu S \\
\dot{E} &= \beta(t) \frac{I}{N} S - (\delta + \mu) E \\
\dot{I} &= \delta E - (\gamma + \mu) I \\
\dot{R} &= \gamma I - \mu R
\end{cases}
\end{equation}

\vspace{-0.5cm}

\paragraph{Incorporating Seasonality in Disease Transmission}

In Bangladesh, dengue transmission is heavily influenced by seasonal variations, particularly during the rainy season, when increased rainfall creates ideal conditions for Aedes aegypti mosquito breeding \cite{morales2016seasonal, pliego2017seasonality}. To account for this seasonality in our model, we introduce a time-dependent transmission rate, $\beta(t)$, which captures the fluctuations in transmission driven by seasonal factors. $\beta(t)$ is modeled as a combination of a baseline transmission rate, $\beta_{np}$ (during non-rainy seasons), and a Gaussian peak transmission rate, $\beta_{p}$ (during rainy seasons), which then decays logistically back to baseline. Additionally, we include $A_{\beta}$ to denote the amplitude of the seasonal variation, $t_{p}$ to represent the peak time at the midpoint of the rainy season, and $\sigma$ to control the width of the seasonal peak. The functional form of $\beta(t)$ is given by:

\begin{equation} \label{eq:beta}
\beta(t) = \beta_{np} + A_{\beta} \exp\left(-\frac{(t - t_{p})^2}{2\sigma^2}\right) + \frac{\beta_{p}}{1 + \exp\left(\frac{t - t_{p} - \sigma}{\sigma}\right)}
\end{equation}

\begin{figure}[!htbp]
    \centering
    \includegraphics[width=0.5\textwidth]{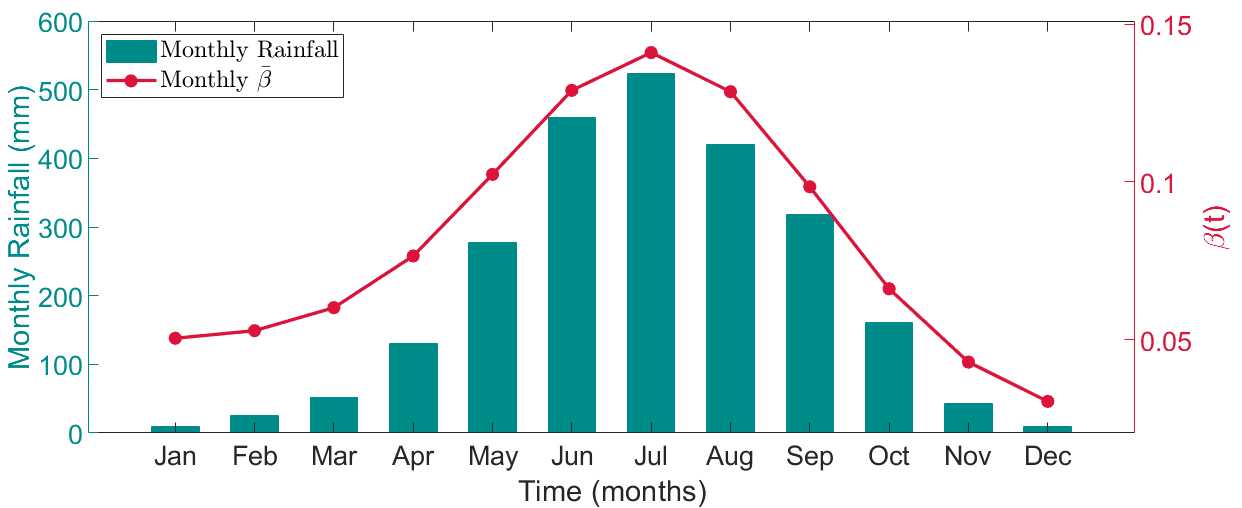}
    \caption{ Time-varying transmission rate $\beta(t)$ and average monthly rainfall in Bangladesh over one year. $\beta(t)$ is evaluated using \eqref{eq:beta} with parameters $A_{\beta}=0.10$, $t_{p}=200$, $\sigma=60$, $\beta_{p}=0.03$, and $\beta_{np}=0.02$. Rainfall data for average monthly levels in Bangladesh in 2023 is collected from \cite{bangladesh_meteorological_department_2023}.}
    \label{fig:beta_rainfall_plot}
\end{figure}

The Gaussian component in  \eqref{eq:beta} models the spike in cases, while the logistic decay accounts for the return to the endemic phase. Incorporating this seasonality allows the model to more accurately reflect the real-world dynamics of dengue transmission in Bangladesh (Fig.~\ref{fig:beta_rainfall_plot}).

\subsection{Parameter Estimation through MCMC}

To estimate the parameters of our model \eqref{eq:odeseir}, we employ a Bayesian inference approach, utilizing the Metropolis-Hastings algorithm within the MCMC framework. 

\paragraph{Priors and Likelihood}
The prior distribution for each parameter \( \theta \) is specified as follows:
\[
\beta_{np} \sim U(0, 1), \quad A_\beta \sim U(0, 1), \quad \beta_p \sim U(0, 1)
\]
\[
t_p \sim N(220, 30), \quad \sigma \sim N(60, 10), \quad \sigma_{obs} \sim \text{Exponential}(1)
\]

The posterior distribution of the parameters, given the observed data \( D \), is obtained using Bayes' theorem:
\[
P(\theta | D) = \frac{P(D | \theta) P(\theta)}{P(D)}
\]
or alternatively:
\[
P(\theta | D) \propto P(D | \theta) P(\theta)
\]
where \( P(D | \theta) \) is the likelihood function and \( P(\theta) \) is the prior.

The likelihood function is constructed by comparing the observed daily dengue cases \( D_i \) to the model-predicted cases \( y(t_i) \). Assuming Gaussian errors, the likelihood is given by:
\[
L(\theta | D) = \prod_{i=1}^{T} \frac{1}{\sqrt{2 \pi \sigma_{obs}^2}} \exp\left( -\frac{(D_i - y(t_i))^2}{2\sigma_{obs}^2} \right)
\]

\paragraph{Model Predictions and Error Handling}
The prediction error \( E^2 \) between the observed and predicted values is minimized during the MCMC sampling process,
\[
E^2 = \sum_{i=1}^{T} (D_i - y(t_i))^2
\]
Assuming the errors are Gaussian, the likelihood function becomes $L(\theta | D) = \exp(-E^2)$.

\paragraph{MCMC Sampling Details}

We run four independent MCMC chains, each with 50,000 iterations and 5,000 warm-up iterations. Convergence is assessed using the Gelman-Rubin statistic \( \hat{R} \), ensuring that the chains have mixed well and the variance between chains is similar to the variance within chains. The estimated parameters are presented  in Table~\ref{tab:seir_posterior_summary_new}, while the results of the posterior predictive check are illustrated in Figure~\ref{fig:posterior}.

\subsection{Time-Nonhomogeneous Stochastic Modeling}

The CTMC-based SEIR model builds on the ODE-based model framework to formulate the infinitesimal transition probabilities for a time-nonhomogeneous stochastic process. For simplicity, the same notation that is used in the ODE model is applied to the CTMC model. The variables in the CTMC are discrete random variables where $E=0$ and $I=0$ are absorbing states: $S(t), E(t), I(t), R(t) \in {0,1,2,3,\dots,N}, \quad t \in [0,\infty)$. Let \(\Delta t\!>\!0\) be sufficiently small so that at most one event occurs during the interval $\Delta t$. Define $\mathbf{X}(t) = (S(t), E(t), I(t), R(t))$ and $\Delta \mathbf{X}(t) = (\Delta S(t), \Delta E(t), \Delta I(t), \Delta R(t))$, where, for example, \(\Delta E(t) = E(t + \Delta t) - E(t)\), and similarly for the other states. The transition probabilities between states depend on both the current time $t$ and the elapsed time between events $\Delta t$, with transition rates summarized in Table~\ref{tab:seirctmc}. For instance, the probability of a new infection occurring during event 4 in Table~\ref{tab:seirctmc} within the time interval \(\Delta t\) is given by: $\mathbb{P}\{\Delta \mathbf{X}(t) = (-1, 1, 0, 0) \mid \mathbf{X}(t)\} = \beta(t) I \frac{S}{N} \Delta t + o(\Delta t)$

\begin{table}[!htbp]
\caption{Transitions and rates in the  CTMC-based model\label{tab:seirctmc}}
\centering
\begin{tabular}{|c|l|l|c|}
\hline
{\bf Event} & {\bf Description} & {\bf State transition} & {\bf Rate }\\ \hline
1 & Susceptible birth & $S \to S+1$ & $\Lambda$ \\ 
2 & Susceptible death & $S \to S-1$ & $\mu S $  \\ 
3 & Recovered death & $R\to R-1$ & $ \mu R$ \\ 
4 & Infection & $(S,E) \to (S-1,E+1)$ & $\beta(t) I\frac{S}{N} $ \\ 
5 & Exposed death & $E\to E-1$ & $ \mu E $ \\
6 & Latent to infectious & $(E,I)\to(E-1,I+1)$ & $ \delta E $ \\ 
7 & Infected death & $I\to I-1$ & $ \mu I $ \\
8 & Infected recovery & $(I,R)\to (I-1,R+1)$ & $ \gamma I  $ \\ \hline
\end{tabular}
\end{table}

\subsection{Model Analysis}

\paragraph{Basic and Seasonal Reproduction Numbers}

When  model parameters are constant, the basic reproduction number, $\mathcal{R}_0$, can be directly computed from the next generation matrix (NGM). This is done by analyzing the exposed and infectious stages of the system \eqref{eq:odeseir} through linearization around the Disease-Free Equilibrium (DFE), where $S = \Lambda/\mu = N$. The DFE remains stable in the absence of infection. In the NGM approach, the Jacobian matrix is decomposed into new infections, $F(t)$, and other transitions, $V(t)$, such that $J(t) = F(t) - V(t)$ \cite{van2002reproduction, wang2008threshold}:

\begin{equation*}
F(t) = \begin{bmatrix} 0 & \beta(t)\\ 0 & 0\end{bmatrix}, \quad V(t) = \begin{bmatrix} \delta + \mu & 0\\ -\delta & \gamma + \mu\end{bmatrix}.
\end{equation*}

With a constant transmission rate, $\beta(t) \equiv \bar{\beta}$, the basic reproduction number, $\mathcal{R}_0$, is given by

\begin{equation}\label{eq:seirR0}
\mathcal{R}_0 = \rho(FV^{-1}) = \frac{\bar{\beta} \delta}{(\delta + \mu)(\gamma + \mu)}.
\end{equation}

When $\mathcal{R}_0 >1$, the DFE is unstable, leading to an outbreak. If  $\mathcal{R}_0 <1$, the DFE is stable, and the disease dies out.

For systems with time-periodic transmission rates, Wang and Zhao’s approach using Floquet theory \cite{klausmeier2008floquet,wang2008threshold} provides a more accurate representation of the seasonal reproduction number ($\mathcal{R}_0^{\text{seasonal}}$). For $X=(E,I)^T$, the linear system can be expressed as

\begin{equation*}
\dot{X} = \left[F(t) - V(t)\right]X, \quad X(0) = I_d.
\end{equation*}

The maximum Floquet multiplier of the monodromy matrix, \(\rho(X(\omega))\), corresponds to the basic reproduction number, \(\mathcal{R}_0\). Since an explicit solution for \(\mathcal{R}_0\) is generally unavailable, it is computed numerically by introducing a parameter \(\lambda\) into the system:

\begin{equation}\label{eq:floquet}
\dot{X} = \left[\dfrac{F(t)}{\lambda} - V(t)\right]X, \quad X(0) = I_d.
\end{equation}

An iterative sequence \(\{\lambda_i\}\) is generated until it converges to a value \(\lambda_k\) such that the spectral radius of the maximum Floquet multiplier is close to one, i.e., \(|\rho(X(\omega, \lambda_k)) - 1| < 10^{-6}\). The final value \(\lambda_k\) approximates $\mathcal{R}_0^{\text{seasonal}}$. The stability of the DFE  and the role of the constant and seasonal basic reproduction numbers as thresholds for disease outbreak are validated by verifying specific conditions \cite{van2002reproduction, wang2008threshold}.

\paragraph{Probability of Disease Outbreak}

To estimate the probability of a disease outbreak from the CTMC model, $n$ sample paths are simulated using a Monte Carlo approach to account for the stochastic nature of disease transmission and progression. Each path begins with 1 initial infected individual, and the simulation runs until one of two conditions is met: either the combined number of exposed and infected individuals ($E+I$) drops to 0, indicating disease extinction, or it reaches a predefined outbreak level (OL). The outbreak probability is calculated as the proportion of sample paths that lead to an outbreak, while the extinction probability is the proportion of paths that lead to extinction. If $n_{\text{ext}}$ represents the number of sample paths where extinction occurs, the extinction probability is given by $\mathbb{P}_{\text{ext}} ={n_{\text{ext}}}/{n}$, and the outbreak probability is defined as $\mathbb{P}_{\text{outbreak}} = 1 - \mathbb{P}_{\text{ext}}$.

\section{Numerical Results}
\label{sec:Numerical Results}
To ensure the reliability of our MCMC sampling, we assess convergence using the Gelman-Rubin diagnostic, also known as the \( \hat{R} \) statistic, and calculated the Effective Sample Size (ESS) for each parameter.

The \( \hat{R} \) values for all parameters are consistently close to 1.0, indicating proper convergence. In terms of the Effective Sample Size (ESS), all parameters has sufficiently large values, confirming that the number of independent samples is adequate for precise estimation of the posterior distributions. The ESS values are as follows: \( A_\beta\)  (47,560), \( \beta_{np}\) (39,800), \( \sigma \) (40,670), \( \beta_p\) (40630), and \( t_{p} \) (77,290). 

\begin{figure}[!htbp]
    \centering
    \includegraphics[width=0.4\textwidth]{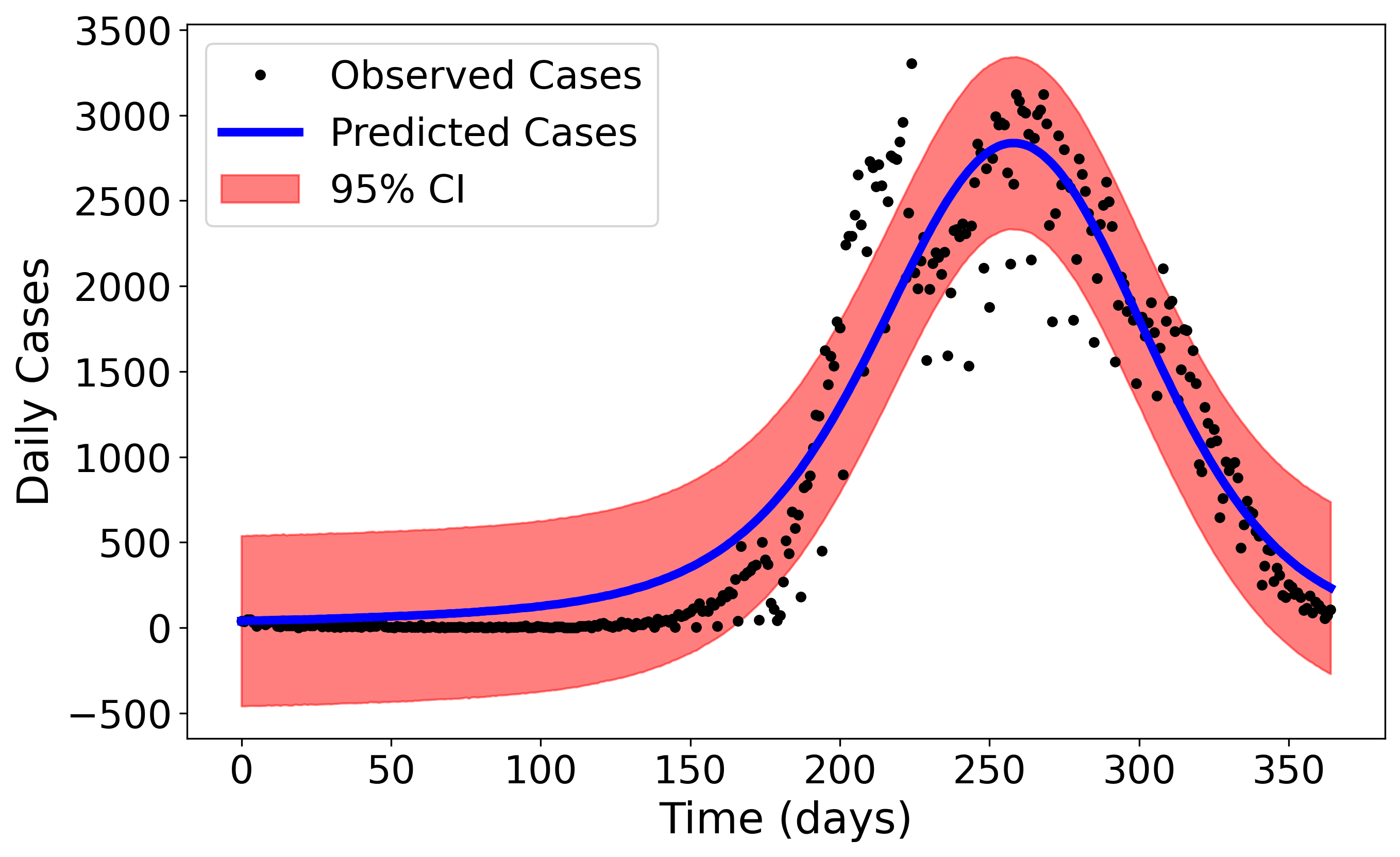}
    \caption{ Posterior predictive check compares the observed daily dengue cases (black points) with the model's predicted cases (blue line) along with the 95\% credible interval (red shaded region).}
    \label{fig:posterior}
\end{figure}

 Fig.~\ref{fig:posterior} shows that the predicted values closely follow the observed data, especially around the peak of the dengue outbreak, indicating a good fit of the model. The 95\% credible interval effectively captures the variability in the observed data, providing a measure of uncertainty around the predictions.

The results presented in Table \ref{tab:seir_posterior_summary_new} summarize the posterior estimates of key parameters in the model. The non-peak transmission rate (\(\beta_{np}\)) is estimated with a mean value of 0.0566 and a 95\% credible interval (CI) ranging from 0.0333 to 0.0741, which is consistent with literature \cite{pandey2013comparing} . The amplitude of seasonal variation (\(A_{\beta}\)) has a mean of 0.0470 and a 95\% CI between 0.0415 and 0.0545, indicating moderate seasonal effects. The peak time (\(t_p\)) for dengue transmission is centered around 190.72 days, which corresponds to early July, aligning with the onset of the rainy season in Bangladesh, a critical period for increased mosquito breeding and dengue transmission.

\vspace{-0.5cm}
\begin{table}[!htbp]
\centering
\caption{Posterior Parameter Estimates for the SEIR Model}
\begin{tabular}{|c|c|c|c|}
\hline
{\bf } & {\bf Prior} & {\bf Mean} & {\bf 95\% CI} \\
\hline
$\beta_{np}$ & $U(0, 1)$ & 0.0567 & [0.033, 0.074] \\
$A_{\beta}$  & $U(0, 1)$ & 0.047 & [0.042, 0.055] \\
$t_{p}$  & $N(220, 30)$ & 190.72 & [190.02, 192.64] \\
$\sigma$ & $N(60, 10)$ & 61.817 & [50.896, 75.117] \\
$\beta_{p}$ & $U(0, 1)$ & 0.085 & [0.069, 0.105] \\
\hline
$\delta$ & -- & $0.25$ \cite{pandey2013comparing} & -- \\
$\gamma$  & -- & $0.125$ \cite{pandey2013comparing} & -- \\
$\mu$  & -- & $0.0000380$ \cite{who_bangladesh_data}
 & -- \\
\hline
\end{tabular}
\label{tab:seir_posterior_summary_new}
\end{table}

\begin{figure}[!htbp]
\centering
\includegraphics[width=\linewidth]{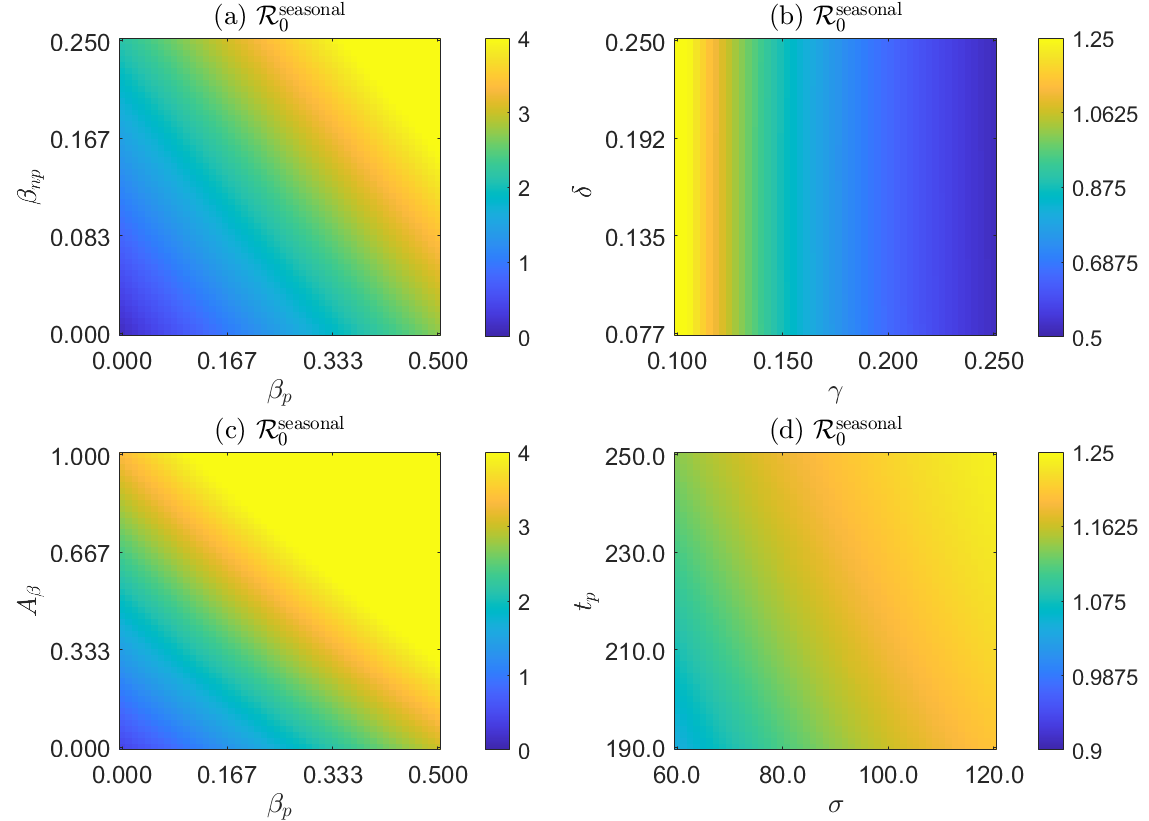} 
\caption{\label{fig:R0_heatmap} Impact of parameter variations on $\mathcal{R}_0^{\text{seasonal}}$ across four cases, with all other parameters held at their respective mean values from Table \ref{tab:seir_posterior_summary_new}.}
\end{figure}

\begin{figure}[!htbp]
\centering
\includegraphics[width=0.5\linewidth]{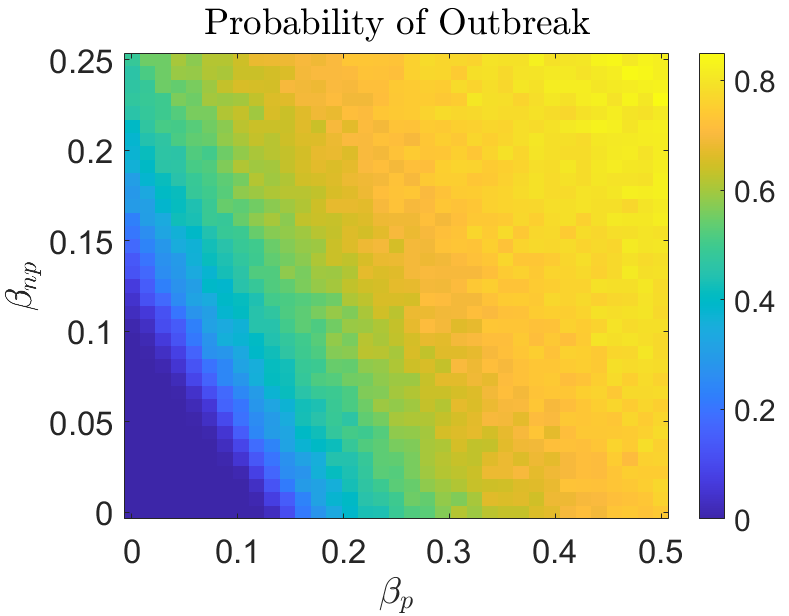} 
\caption{\label{fig:P_outbreak} Impact of varying $\beta_{np}$ and $\beta_p$ on the probability of outbreak ($\mathbb{P}_{\text{outbreak}}$), with all other parameters held at their respective mean values from Table \ref{tab:seir_posterior_summary_new}, $n=1000$, and $OL=100$. }
\end{figure}

In Fig.~\ref{fig:R0_heatmap}, the heatmaps illustrate how changes in model parameters affect the $\mathcal{R}_0^{\text{seasonal}}$. For a comprehensive analysis, we extend the model parameter values beyond the 95\% confidence interval, but remain well within the range found in existing literature \cite{pandey2013comparing}. Furthermore,  Fig.~\ref{fig:P_outbreak} demonstrates that varying the same parameters as in  Fig.~\ref{fig:R0_heatmap} (a) produces a similar pattern in the probability of outbreak heatmap, highlighting how our framework can generate consistent and comparable results from both deterministic and stochastic models.

\vspace{-0.5cm}

\begin{table}[!htbp]
    \centering
       \caption{Comparison between basic $\mathcal{R}_0$, $\mathcal{R}_0^{\text{seasonal}}$, and $\mathbb{P}_{\text{outbreak}}$  as $\sigma$,$\beta_{np}$, and $\beta_{p}$ are varied, with $n=10000$, and $OL=100$}
    \label{tab:R0_pext_table}
     \begin{tabular}{|c|c|c|c|c|c|c|}
       \hline
        \textbf{$\sigma$} & \textbf{$\beta_{np}$} & \textbf{$\beta_{p}$} & \textbf{$\bar{\beta}$}& \textbf{$\mathcal{R}_0$} & \textbf{$\mathcal{R}_0^{\text{seasonal}}$} & \textbf{$\mathbb{P}_{\text{outbreak}}$} \\
        \hline
        60 days & 0.057 & 0.085 & 0.132 & 1.059 & 1.045 & 0.113 \\
        75 days & 0.057 & 0.085 & 0.139 & 1.108 & 1.097 & 0.137 \\
        120 days & 0.057 & 0.085 & 0.151 & 1.205 & 1.202 & 0.185\\
         \hline
        60 days & 0.074 & 0.105 & 0.163 & 1.306 & 1.290 & 0.300 \\
        75 days & 0.074 & 0.105 & 0.170 & 1.358 & 1.347 & 0.309\\
        120 days & 0.074 & 0.105 & 0.183 & 1.460 & 1.456 & 0.330\\
         \hline
        60 days & 0.057 & 0.581 & 0.463 & 3.704 & 3.578 & 0.765 \\
        75 days & 0.057 & 0.581 & 0.479 & 3.827 & 3.743 & 0.784\\
        120 days & 0.057 & 0.581 & 0.505 & 4.036 & 4.003 & 0.809\\
     \hline
    \end{tabular}
 
\end{table}

\vspace{0.5cm}
\section{Conclusion}
\label{sec:Conclusion}

This research aims to equip policymakers with a comprehensive mathematical framework to better understand disease dynamics and address the growing dengue crisis in Bangladesh. We incorporate the evident seasonality of dengue outbreak patterns into our model and propose an innovative transmission rate that reflects these seasonal variations. Our study demonstrates how Bayesian parameter estimation can guide both deterministic and stochastic SEIR frameworks, allowing for the comparison of key outbreak measures like basic and seasonal $\mathcal{R}_0$ and probability of outbreak, as shown in Table \ref{tab:R0_pext_table}, ultimately aiding in the mitigation of dengue infections.

\end{document}